\documentclass{evn2004}
\usepackage{txfonts}

\begin{document}
\setcounter{page}{269}

   \title{Cryogenic Filters for RFI Mitigation in Radioastronomy}

   \author{G. Tuccari\inst{1}\and
           A. Caddemi\inst{2}\and
	     S. Barbarino\inst{3}\and
	     G. Nicotra\inst{1}\and
           F. Consoli\inst{3}\and
           F. Schillir\'o\inst{1}\and
           F. Catalfamo\inst{2}
          }

   \institute
{Radioastronomy Institute CNR - Noto Section, Ctr.da Renna, 96017 Noto (Sr), Italy\and
Dipartimento di Fisica della Materia e Tecnologie Fisiche Avanzate e INFM, Messina University, Italy\and
Dipartimento di Fisica e Astronomia, Catania University, Italy
}

   \abstract{
RFI mitigation in Radioastronomy can be achieved adopting cryogenic filters 
in appropriate typologies. A study has been conducted in L, C and X band with 
the evaluation of the filter architecture in copper, with theoretical estimation, 
computer simulations, prototypes realization, laboratory measurements. 
Such work has been preliminary to the realization of HTS samples with the purpose 
of a similar complete characterization approach.
   }

   \maketitle
%

\section{Introduction}
Interferometric observations in radioastronomy and in particular in VLBI 
are less sensitive to the RFI presence with respect to a single dish total 
power system. Indeed an interferometer primarily responds to correlated 
signals, then radio interference present at only one telescope produces 
reduced effects, that are strongly attenuated as long as the integration 
time increases. Nevertheless strong signals can affect the general 
interferometer performance because of the inter-modulation products. 
An other effect the interferometer is affected by is the de-correlation 
of the bandwidth in case of wide band interference as fringe rate at 
opposite edges of the observed band is in general much different. 
Moreover any additive contribution to the receiver noise acts as decreasing 
factor for the correlation coefficient.\\

A good approach to properly take into account the RFI effects this is adopting 
general rules to set as high as possible the spurious free dynamic range 
increasing the receiver OIP3 factor and accurately studying the nature of the 
observed band in order to better adapt the receiver with the environment where 
it has to operate. While the first is in some way straightforward to implement, 
the second requires an accurate determination of the interfering signals in the 
band and outside, because harmonics could introduce products in the observed band. 
Different solutions are possible and the filtering option is the most obvious, 
while it is not too obvious where to insert a filter and which type is to be selected: 
the choice indeed depends on the specific RFI spectrum occupancy. A band-pass 
filter can be inserted in the LNA amplification either as an inter-stage filter 
or as front-end. Such filter insertion can be realized inside the LNA enclosure 
and depending on the position the realization technology has to be chosen between 
Cu and superconductiong materials.\\

The different solutions have as common factor the need to take into account that 
the LNA operates, for adding a reduced noise contribution, at cryogenic temperatures. 
A filter inserted in the cryogenic path should then behave noise performance minimal 
enough to keep its insertion still valuable. While this contribution could be 
important in an intermediate stage, it is an essential element when the filter is 
the first component of the chain. A trade-off evaluation has to be performed 
determining the conditions when the increased noise temperature due to the filter 
insertion is worth with respect to the reduced noise contribution due to the 
interfering signal.


\section{Filters performance}

An analysis has then been undertaken to determine the filter typologies that 
better fits with their insertion in a LNA enclosure, then considering linear 
dimension and area occupancy. Pass-band filters have been examined as inter-stage 
elements planning their realization with the same substrates and adopting materials 
and technologies used for the low noise amplifier realization. In the case of front-end 
filters, HTS (High Temperature Super-conducting) materials have been taken into consideration. 
For the sake of simplicity and practical realization, lumped element filters have not 
been examined, but only microstrip typologies. More software packages have been used and 
compared to simulate the performance and in some cases the filter realization has been 
performed to verify the simulation goodness and actual filter performance. 
Such prototypes have even been measured at cryogenic temperature to evaluate their 
effective impact on the overall performance. Few architectures have been proved particularly 
valuable for the goal to achieve. In some cases a design for the HTS version has been produced 
and the methods to implement such kind of filters are under an advanced detailed study; 
the first prototype will be shortly measured in L band.\\

Studies have been done in C band in the range 4.70-5.05, band commonly adopted in VLBI 
radioastronomy, taking into consideration square open-loop typology in quasi-elliptic 
structures. Simulations have been realized with some practical realization for comparison. 
Adopting Cu on RT/Duroid6010LM, at 300 K temperature behaving er=10.2, with 1.7dB typical 
attenuation in band and a minimum rejection of -32.4 dB for frequency values in the lower 
side, while -20.4 in the upper side, with a typical selectivity of 115 dB/GHz. The same 
filter at 77 K, having an er=11.75, presents a mean attenuation in band of 1.2 dB with 
similar minimum rejection values while a selectivity of 128 dB/GHz. This project has been 
transferred to an HTS material YBCO on LaAlO3, with er=23.6 operating at 77 K, and showing 
an attenuation of 0.1 dB in band, minimal rejection of -32.5 dB and -25.2 dB, with the 
interesting selectivity values of 215 dB/GHz and 362 dB/GHz in the lower and upper side of 
the filter frequency pass-band.\\

In the X band for the frequency range 8.18 - 8.58 GHz, still widely adopted in VLBI, similar 
studies reported at 300 K for Cu on Duroid a typical attenuation in band of 2.5 dB with minimal 
rejection of 37.1 dB and 22.8 dB for lower and upper band sides respectively. Similar rejection 
values have been found for the HTS material, with a much reduced attenuation in band of 0.06 dB.\\

Linear phase filter shave even been analyzed trying to optimize the phase linearity in 
the range of the L band 1.40 - 1.72 GHz, adopting Cu on Duroid at 300 K. Attenuation of 
0.4-1.2 dB in band is obtained for a group delay of 1.4 - 1.9 ns. This kind of project 
showed particular difficulties for taking under reasonable values all the fundamental 
parameters.\\

Finally an HTS filter in L band range 1.40 - 1.72 GHz has extensively studied to produce 
a real sample to be adopted for a radar emission mitigation. Typology considered was Hairpin 
with YBCO on LaAlO3. Simulated performance show still pretty good attenuation values in band, 
less than 0.1 dB, return loss better than -20 dB, minimal rejection better than -40 dB and 
-30 dB respectively for lower and upper sides of the pass-band, while slopes are 165 dB/GHz 
and 160 dB/GHz. The realization of the filter prototype is at time of writing in advanced 
phase and it will be reported.
%

\section{Conclusions}

Integrating filters inside LNA structures is an interesting option in order to 
mitigate effects of severe RFI. A good method is to determine the spectrum to 
observe inserting pass-band filters at a stage where the best compromise is found 
between benefits and drawbacks. Cryogenic temperature brings moderate benefits on Cu lines, 
while much better performance are obtained with superconducting materials at the 
expenses of increased cost and more difficult implementation. It's anyway valuable 
in some cases adopt Cu structures in intermediate stages. 
The trend we think useful to follow is to gain experience components so to adopt 
them when the RFI  situation suggest their use. So much effort is worth to be spent on 
this subject for an careful costs versus benefits evaluation.

   \begin{figure}
   \centering
   \vspace{200pt}
   \includegraphics{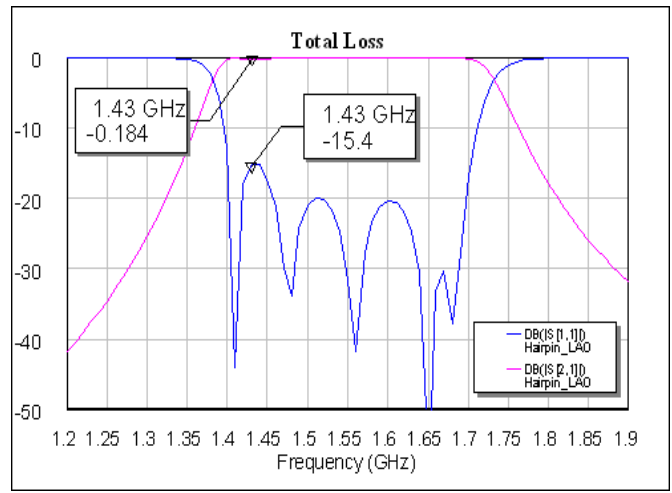}
   \caption{HTS filter response simulation in L band.}
   \end{figure}

\end{document}